\documentclass[prl,twocolumn,showpacs,preprintnumbers,superscriptaddress]{revtex4}
\usepackage{graphicx}
\usepackage{color}
\usepackage{dcolumn}
\usepackage{bm}
\usepackage{graphics}

\usepackage{multirow}
\usepackage{hhline}
\usepackage{subfig}
\usepackage{amsbsy}
\usepackage{amsmath,amssymb}
\usepackage{amsfonts}
\usepackage{amsthm}
\usepackage{float}
\newcommand{\ket}[1]{| #1 \rangle}

\theoremstyle{plain}

\theoremstyle{definition}

\begin{document}

\title{Retrieving and Routing Quantum Information in a Quantum Network}
\author {S. Sazim}
\affiliation{Institute of Physics, Sainik School Post, 
Bhubaneswar, Orissa, 751005,  India.}
\author {V. Chiranjeevi}
\affiliation{International Institute of Information Technology, Gachibowli, Hyderabad 500 032, Telangana, India.}
\author {I. Chakrabarty}
\affiliation{International Institute of Information Technology, Gachibowli, Hyderabad 500 032, Telangana, India.}
\author {K. Srinathan}
\affiliation{International Institute of Information Technology, Gachibowli, Hyderabad 500 032, Telangana, India.}
\begin{abstract}
In extant quantum secret sharing protocols, once the secret is shared in a quantum network (\textsc{qnet}) it cannot be retrieved, even if the dealer wishes that his/her secret no longer be available in the network. For instance, if the dealer is part of two \textsc{qnet}s, say $\mathcal{Q}_1$ and $\mathcal{Q}_2$ and he/she subsequently finds that $\mathcal{Q}_2$ is more reliable than $\mathcal{Q}_1$, he/she may wish to transfer all her secrets from $\mathcal{Q}_1$ to $\mathcal{Q}_2$. Known protocols are inadequate to address such a revocation. In this work we address this problem by designing a protocol that enables the source/dealer to bring back the information shared in the network, if desired. Unlike classical revocation, the no-cloning theorem automatically ensures that the secret is no longer shared in the network.

The implications of our results are multi-fold. One interesting implication of our technique is the possibility of {\em routing}\/ qubits in  {\em asynchronous} \textsc{qnets}. By asynchrony we mean that the requisite data/resources are intermittently available (but not necessarily simultaneously) in the \textsc{qnet}.  For example, we show that a source $S$ can send quantum information to a destination $R$ even though (a) $S$ and $R$ share no quantum resource, (b) $R$'s identity is {\em unknown}\/ to $S$ at the time of sending the message, but is subsequently decided, (c) $S$ herself can be $R$ at a later date and/or in a different location to bequeath her information (`backed-up' in the \textsc{qnet}) and (d) importantly, the path chosen for routing the secret may hit a dead end due to resource constraints, congestion, etc., (therefore the information needs to be {\em back-tracked} and sent along an alternate path). Another implication of our technique is the possibility of using {\em insecure}\/ resources. For instance, if the quantum memory within an organization is insufficient, it may safely store (using our protocol) its private information with a neighboring organization without (a) revealing critical data to the host and (b) losing control over retrieving the data. 
 
Putting the two implications together, namely routing and secure storage, it is possible to envision applications like quantum mail (qmail) as an outsourced service.
\end{abstract}

\maketitle

Quantum entanglement \cite{einstein} not only gives us insight into understanding the deepest nature of reality but also acts as a very useful resource in carrying out various information
processing protocols like quantum teleportation \cite{bennett2, Mattle}, 
quantum cryptography \cite{gisin} and quantum secret sharing(QSS) \cite{hillery}, to name a few.\vskip 0.1cm 

In a secret sharing protocol the sender/dealer of the secret message, who is unaware of the individual honesty of the receivers, shares the secret in such a way that none of the receivers get any information about the secret. QSS \cite{hillery, cleve99} deals with the problem of sharing of both classical and quantum secrets. A typical protocol for quantum secret sharing, like many other tasks in quantum cryptography, uses entanglement  as a cardinal resource, mostly pure entangled states. Karlsson et al. \cite{karlsson} studied QSS protocols using bipartite pure entangled states as resources. Many authors investigated the concept of QSS using tripartite pure entangled states and  multi-partite states like graph states \cite{bandyopadhyay,bagherinezhad,lance,gordon,zheng, markham,markham08}. Li et al. \cite{li} proposed semi-quantum secret sharing protocols taking maximally entangled GHZ state as resource.

In a realistic situation, the secret sharing of classical or quantum information involves transmission of qubits through noisy channels that entails mixed states. Recently in \cite{satya}, it is shown that QSS is possible with bipartite two-qubit mixed states (formed due to noisy environment or otherwise). Subsequently in \cite{indranil} authors propose a protocol for secret sharing of classical information with three-qubit mixed state. Quantum secret sharing has also been realized in experiments \cite{tittel, schmid, schmid1, bogdanski}.

In QSS, it is typically assumed that the system consists of solely the dealer and the receivers. However, in practical settings the dealer/receivers are part of a quantum network. One important question of how information can be transferred through a quantum network is addressed in \cite{ind, delgado, Paparo, lemr, klemr}. In this work we focus on two different situations in a given quantum network (\textsc{qnet}). 
In the first situation, we consider the problem of revoking the secret in QSS. For instance, if the dealer finds the receivers to be dishonest, she can stop them from accessing it. Moreover, she may choose to retrieve the secret completely. In our model we consider the receivers to be semi-honest -- that is, the receivers, though dishonest to eavesdrop on their share and process it, diligently participate in the protocol. On the other hand, note that Byzantinely malicious receivers can easily destroy the secret, making revocation impossible. 
In the second situation we have extended the above idea to design routing mechanism for multi-hop transmission of {\em secret}\/ qubits in the shared domain itself.

Although the above two situations appear to solve unrelated problems namely, revocable secret sharing and quantum routing, the following is an interesting symbiosis of the two to solve problems posed by resource constraints and asynchrony in the network. Consider a situation where quantum storage is constrained and therefore Alice needs to store her private data in some (probably untrustworthy) memory available in the network. This she can do using {\em  revocable}\/ QSS. Further, if she wants to send these data to Bob (for security reasons) she should be able to do it without reconstructing the quantum secret anywhere in the network. This she can achieve using the quantum routing in shared domain. Incidentally, our solution also takes care of scenarios where Bob too is in short supply of trusted quantum memory and uses network storage.\\ \\

\noindent\textit{\bf Sharing of a Message :}\\
First of all, we consider a simple situation where we have three parties Alice, Bob and Charlie. 
They share a three-qubit maximally entangled GHZ state, i.e., $\ket{GHZ}_{ABC}=\frac{1}{\sqrt{2}}(\ket{000}+\ket{111})$. 
Here the first qubit is with Alice, second is with Bob and the third one is with Charlie. 
Here Alice is the dealer and she wishes to secret-share a qubit $\ket{S}=\alpha\ket{0}+\beta\ket{1}$ (where $\lvert\alpha\rvert^2+\lvert\beta\rvert^2=1$; 
$\alpha,\beta$ are amplitudes) with both the parties Bob and Charlie. 
 In order to do so Alice has to do two-qubit measurements in Bell basis 
$\{ |\phi_{\pm}\rangle,|\psi_{\pm}\rangle\}$ jointly on 
her resource qubit and the message qubit she wants to share (see Appendix 1). 
In correspondence with various measurement outcomes obtained by Alice, Bob and Charlie's qubits collapse into the states given in TABLE 1.
\begin{table}[h]
\caption{\bf Sharing of Quantum Information}
\begin{center}
\begin{tabular}{|c|c|}
\hline \parbox{1.25in}{\ \\Alice's Measurement Outcomes\\ \ \\} & \parbox{1.75in}{\ \\Bob and Charlie's Combined State\ \\ \ \\}\\
\hline $\ket{\phi^+}$  & $\alpha\ket{00}+\beta\ket{11}$  \\
\hline $\ket{\phi^-}$ & $\alpha\ket{00}-\beta\ket{11}$  \\
\hline $\ket{\psi^+}$ & $\alpha\ket{11}+\beta\ket{00}$   \\
\hline $\ket{\psi^-}$ & $\alpha\ket{11}-\beta\ket{00}$  \\
\hline
\end{tabular} 
\end{center}
\end{table}

At this point if Alice finds both Bob and Charlie to be dishonest, she can stop them from accessing the message. She does this by not communicating about her measurement results to any one of them. So there is  
no transfer of classical bits at this stage. At this point there lies the question of security from Bob and Charlie sides. If we have malicious (parties who are not going to follow the protocol and do whatever they wish to do) Bob and Charlie can destroy the message by doing local operations in their respective qubits and by communicating classically between them. 
However, they will never be successful in obtaining the message without Alice's help.\\ 

\noindent\textit{\bf Revocation of Quantum Information :}

If Bob and Charlie are semi-honest (i.e., they are faithful executors of the protocol but curious to learn Alice's secret), we ask {\em can Alice revoke her shared secret $\ket{S}$}? The ability to revoke the shared secret is important for several reasons, some of which are (a) Alice decides to change her secret (for instance, $\ket{S}$ might have been inadvertently shared), (b) Alice conjectures that the recipients are no longer trustworthy, (c) there is an update of data/secrets in the higher-level application using secret sharing as a subroutine and (d) Alice has found a more economical alternative \textsc{qnet} to safeguard $\ket{S}$\\
To make the revocation possible Alice needs an additional resource (a Bell state) shared with Bob. Consider a very simple case when Alice and Bob are sharing the Bell state $|Bell\rangle_{AB}=\frac{1}{\sqrt{2}}(|00\rangle+|11\rangle)$ in addition to the GHZ state shared by Alice, Bob and Charlie. 
Let us also assume the first case in the above TABLE 1, when Bob and Charlie share the entangled state $\alpha|00\rangle+\beta|11\rangle$ 
as a result of Alice's measurement. 
Now Alice asks Bob to do Bell measurement on his two qubits (one from the shared resource and one from shared secret) and Charlie to do  measurement (on his qubit of shared secret) in Hadamard basis (see Appendix 2, see FIG. 1). In TABLE 2 we show how Alice can retrieve back her message by enlisting down the respective local operations corresponding to Bob's and Charlie's measurement outcomes. It can be observed that the Revocation needs Alice to share an additional resource (a Bell state) with at least one of Bob or Charlie. In case if Alice shares a Bell state with Charlie, Charlie will first do Bell measurement on his  two qubits and then Bob will do measurement in Hadamard basis on his qubit. In either case Alice can retrieve her message by performing the respective local operations.
\begin{figure}[h]
\begin{center}

\[
\begin{array}{cc}
\includegraphics[width=0.4\textwidth]{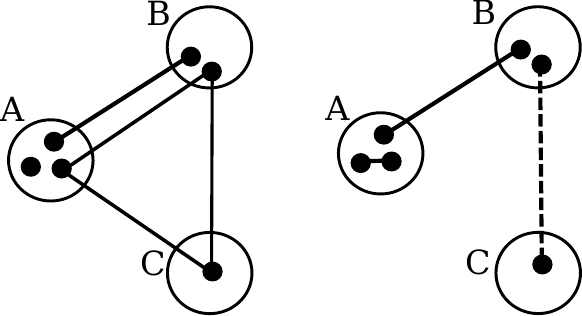}
\end{array}
\]
\caption{\small{\noindent Figure on the left side indicates a three-qubit GHZ state (depicted by a triangle) shared among Alice(A), Bob(B) and Charlie(C) and a two-qubit Bell state shared between and Alice(A) and Bob(B). Alice is also having the secret (depicted by an isolated dot) with herself. The figure on the right describes the situation after Alice's measurement, where both Bob and Charlie are sharing the secret between them (the dotted line).}}\label{SS1}
\end{center}
\end{figure}

\begin{table}[h]
\caption{\bf Retrieving Quantum Information}
\begin{center}
\begin{tabular}{|c|c|c|c|}
\hline \parbox{.75in}{\ \\Bob's  Outcomes\\ \ \\} & \parbox{.75in}{\ \\Charlie's Outcomes\\ \ \\} &\parbox{.75in}{\ \\Alice's Resultant state\\ \ \\} & \parbox{.75in}{\ \\Alice's Local Operations\ \\ \ \\}\\
\hline $\ket{\phi^{+}}$  & $\ket{+}$ & $\alpha|0\rangle+\beta|1\rangle$ & $I$    \\
\hline $\ket{\phi^{+}}$  & $\ket{-}$ & $\alpha|0\rangle-\beta|1\rangle$ & $\sigma_z$  \\
\hline $\ket{\phi^{-}}$  & $\ket{+}$ & $\alpha|0\rangle-\beta|1\rangle$ & $\sigma_z$  \\
\hline $\ket{\phi^{-}}$  & $\ket{-}$ & $\alpha|0\rangle+\beta|1\rangle$ & $I$ \\
\hline $\ket{\psi^{+}}$  & $\ket{+}$ & $\beta|0\rangle+\alpha|1\rangle$ & $\sigma_x$    \\
\hline $\ket{\psi^{+}}$  & $\ket{-}$ & $\alpha|1\rangle-\beta|0\rangle$ & $\sigma_x\sigma_z$  \\
\hline $\ket{\psi^{-}}$  & $\ket{+}$ & $\beta|0\rangle-\alpha|1\rangle$ & $\sigma_x\sigma_z$    \\
\hline $\ket{\psi^{-}}$  & $\ket{-}$ & $\alpha|1\rangle+\beta|0\rangle$ & $\sigma_x$  \\
\hline
\end{tabular} 
\end{center}
\end{table}

For example, suppose Alice shares additional resource with Bob and Bob gets $|\psi^-\rangle$ as outcome of Bell measurement and Charlie gets $|+\rangle$ as outcome of her measurement in Hadamard basis. From Appendix 2, it is clear that Alice will have $\alpha|1\rangle-\beta|0\rangle$ as result of these measurements.

Now, from TABLE 2, Alice needs to apply $\sigma_x\sigma_z$ on her part, i.e., $\sigma_x\sigma_z(\alpha|1\rangle-\beta|0\rangle) = \alpha|0\rangle+\beta|1\rangle$ to complete the revocation.\\ \\

\noindent\textit{\bf Quantum Routing in shared domain :}

If Alice has shared her secret qubit $\ket{S}$ in some part of a (huge) \textsc{qnet}, we ask {\em can she/anyone else retrieve $\ket{S}$ at some other part of the network }? A naive way out is to reconstruct $\ket{S}$ and teleport it, possibly via successive entanglement swapping. However, this severely compromises the security of $\ket{S}$. A superior approach is to retain $\ket{S}$ in the shared domain, while the shares are being routed across the \textsc{qnet}. However, since the shares are themselves entangled and distributed across multiple parties, it is non-trivial to teleport them over the \textsc{qnet}. We address the problem in two parts. First, we show its possible for Alice to dynamically choose the receiver (of her secret), {\em after}\/ the sharing phase. Second, we show that quantum information can be transmitted in the shared domain: that is, the information secret shared among a set of nodes is transferred to another set of nodes. Putting the two together, Alice can now move her shared secret close to the desired receiver in the \textsc{qnet} and also remotely control the reconstruction of the secret at the receiver.

Consider a situation where we have $(3+n)$ parties. Here Alice is the sender, both Charlie and Bob act as agents, the remaining $n$ parties \{$R_1, R_2, R_3,\ldots,R_n$\} are the potential receivers.  Alice desires to send the message in form of a qubit to any one of them. Here the role of Bob and Charlie is changed as they are no longer receivers of information but they now act as agents for holding the information in the network. In broader sense they together act like  a router and play a vital role in sending the information to the desired receiver.\\

Once again we start with Alice, Bob and Charlie sharing a three-qubit 
maximally entangled GHZ state, i.e., $\ket{GHZ}_{ABC}=\frac{1}{\sqrt{2}}(\ket{000}
+\ket{111})$ and Charlie shares  Bell's states, i.e., $\ket{Bell}_{CR_i}=\frac{1}{\sqrt{2}}(\ket{00}+\ket{11})$ with each of the receivers ($R_i$). (In principle, receivers can share resource with any one of the agents Bob and Charlie. Without any loss of generality we assume the receivers share resources with Charlie only.)  
Suppose Alice wishes to send a qubit $\ket{S}=\alpha\ket{0}+\beta\ket{1}$ to $R_i$ through the parties Bob and Charlie. First Alice shares her secret with Bob and Charlie in the same way as it is shown in the (TABLE 1). At this point, Alice sends her measurement outcomes encoded in the form of two classical bits to $R_i$. Once the two bits of classical information are obtained, the receiver can easily get back the Alice's secret $\ket{S}$, provided Bob and Charlie perform the actions as described next. We assume that the identity of the receiver is authentically known to Alice, Bob and Charlie, perhaps through a classically secure authentication/identification protocol.\\

\begin{figure}[h]
\begin{center}

\[
\begin{array}{cc}
\includegraphics[width=0.3\textwidth]{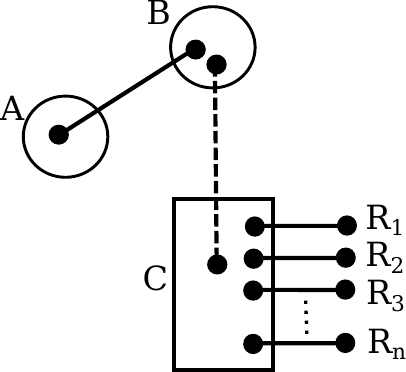}
\end{array}
\]
\caption{\small{ \noindent  Any one of the $n$ receivers \{$R_1, R_2, R_3,\ldots,R_n$\} which individually share Bell states with Charlie can reconstruct the secret.}} \label{router}
\end{center}
\end{figure}

The agents Bob and Charlie do the following. Bob measures his qubit (part of the GHZ state) in the Hadamard basis. Charlie measures two qubits (one from GHZ state and one from Bell state shared with $R_i$) in the Bell basis. After performing these measurements both the agents will send their outcomes through classical channels to the receiver $R_i$. With these measurement outcomes 
the receiver can retrieve the message which Alice intended to send (see Appendix 3, see FIG \ref{router}). Let us consider the case, when Alice and Bob share the 
entangled state $\alpha|00\rangle+\beta|11\rangle$, obtained as a result of Alice's measurement. 
TABLE 3 gives an elaborate view of the unitary operations the receiver $R_i$ has to do upon getting various 
measurement outcomes from Bob and Charlie. 
\begin{table}[h]
\caption{\bf Sending Quantum Information}
\begin{center}
\begin{tabular}{|c|c|c|}
\hline \parbox{1in}{\ \\Charlie's Outcome\\ \ \\} & \parbox{1in}{\ \\Bob's  Outcome\\ \ \\} & \parbox{1in}{\ \\Unitary operations of $R_i$\ \\ \ \\}\\
\hline $\ket{\phi^{+}}$  & $\ket{+}$ & $I$    \\
\hline $\ket{\phi^{-}}$  & $\ket{+}$ & $\sigma_z$  \\
\hline $\ket{\phi^{+}}$ & $\ket{-}$ & $\sigma_z$   \\
\hline $\ket{\phi^{-}}$ & $\ket{-}$ & $I$   \\
\hline $\ket{\psi^{+}}$ & $\ket{+}$ & $\sigma_x$   \\
\hline $\ket{\psi^{-}}$ & $\ket{+}$ & $\sigma_z\sigma_x$\\
\hline $\ket{\psi^{+}}$ & $\ket{-}$ & $\sigma_z\sigma_x$   \\
\hline $\ket{\psi^{-}}$ & $\ket{-}$ & $\sigma_x$   \\
\hline
\end{tabular} 
\end{center}
\end{table}


\begin{figure}[t]
\begin{center}

\[
\begin{array}{cc}
\includegraphics[width=0.35\textwidth]{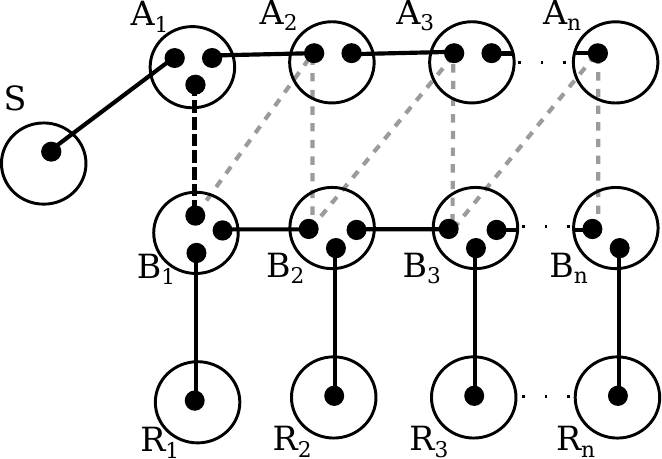}
\end{array}
\]
\caption{\small{A typical quantum mail sending network where $S$ is the source, $(A_i, B_i)$ are the agents and $R_i$ are the receivers. Initially, the information is shared between the pairs $(A_1, B_1)$ (the dotted line) and will be transferred to other pairs until the pair close to the desired receiver is reached. The information is moved along $(A_1, B_1), (B_1, A_2), (A_2, B_2), (B_2, A_3)$ and so on till $(A_n, B_n)$ as shown with gray dotted lines.}}\label{qmaildiag1-eps-converted-to}
\end{center}
\end{figure}
Finally, we address the problem of transferring secret qubits in the shared domain till it comes close to the desired receiver. If we have a source $(S)$ and receivers $R_1, R_2,\ldots, R_n$ and we want to send the information to the receiver $R_i$ through a huge network with pair of agents $(A_1,B_1),(A_2,B_2),\ldots,(A_n,B_n)$ at each blocks, every pair shares Bell state with consecutive pair say $A_i$ with $A_{i+1}$ and $B_i$ with $B_{i+1}$. The above setting is depicted in FIG. 3. Once the source shares the information with $i^{th}$ pair the information can be transferred to $(i+1)^{th}$ pair by the process of entanglement swapping in the following way. $A_i$ performs the Bell measurement on two qubits one from the shared secret and other from the Bell state shared with $A_{i+1}$, similarly $B_i$ performs the Bell measurement on two qubits one from the shared secret and other from the Bell state shared with $B_{i+1}$ (See Appendix 4). This sequence of measurements goes on till the closest pair gets the shared secret. The classical outcomes of each measurement are sent to Alice immediately after the measurement to keep track of the state of the shared secret. The receivers can stay in the network in between each pairs. The source is not going to send the classical information until the quantum information (shared secret) reaches the pair $(A_i,B_i)$ close to the desired receiver. Thus, in a \textsc{qnet} we can share, retrieve, hold and as well as transfer the quantum information.\\

\noindent\textit{\bf Security Analysis:}\\ \\
In this work, we look at QSS as sharing of quantum information with semi-honest participants. By semi-honest, we mean (i) the participants are eager to know any information about the secret without violating the rules of the protocol, (ii) the participants are not allowed to bring ancillary states and perform measurements along with the states involved in the protocol.

For the Revocation protocol, we do security analysis at different stages: (a) Once the secret is shared by protocol given in [4], what is the chance of Bob and/or Charlie getting the secret $|\psi\rangle$? (b) During reconstruction, what is the chance of Charlie getting the secret once Bob has done with his Bell measurement? (c) After the revocation protocol, can Bob and Charlie guess in together what is the secret shared?

Further for reconstruction of the secret at different receiver $R_i$ (FIG. 2) we ask the question (d) if $R_i$ is dishonest, what is the chance that $R_i$ will have the information about the secret after Charlie's measurement.

It is to be noted that the most important part of the protocol is to keep as secret, the classical information of Alice's measurement outcome. In particular, if the classical information is known to other parties having shares, the secret is revealed. Now it remains interesting to see how much information rest of the parties can obtain about the shared quantum state without having access to the classical information. It is intuitive that the information about the shared quantum state is present in the reduced density matrix of the information seekers.
We answer all the above questions by looking at the reduced density operator of the corresponding subsystem and we show that it has no dependence on the actual secret $|\psi\rangle$.

To answer (a), consider the density operator of the system given in Appendix 1 as 
\begin{eqnarray}
\begin{split}
\rho & = \frac{1}{4}\{|\phi^+\rangle\langle\phi^+|(\alpha|00\rangle +\beta|11\rangle)(\alpha^*\langle00| + \beta^*\langle11|)\nonumber\\& \quad 
+ |\phi^-\rangle\langle\phi^-|(\alpha|00\rangle-\beta|11\rangle)(\alpha^*\langle00|-\beta^*\langle11|)\nonumber\\& \quad 
+ |\psi^+\rangle\langle\psi^+|(\alpha|11\rangle+\beta|00\rangle)(\alpha^*\langle11+\beta^*\langle00|)\nonumber\\& \quad 
+ |\psi^-\rangle\langle\psi^-|(\alpha|11\rangle-\beta|00\rangle)(\alpha^*\langle11|-\beta^*\langle00|)\}.
\end{split}
\end{eqnarray}

Tracing out Alice's two-qubit system, the reduced density operator of Bob and Charlie's system is given by,
\begin{eqnarray}
\begin{split}
\rho^{BC} & = \frac{1}{4}\{(\alpha|00\rangle +\beta|11\rangle)(\alpha^*\langle00| + \beta^*\langle11|)\nonumber\\& \quad 
+ (\alpha|00\rangle-\beta|11\rangle)(\alpha^*\langle00|-\beta^*\langle11|)\nonumber\\& \quad 
+ (\alpha|11\rangle+\beta|00\rangle)(\alpha^*\langle11|+\beta^*\langle00|)\nonumber\\& \quad  
+ (\alpha|11\rangle-\beta|00\rangle)(\alpha^*\langle11|-\beta^*\langle00|)\} \\
& =  \frac{1}{2}\{|00\rangle\langle00| + |11\rangle\langle11|\}.
\end{split}
\end{eqnarray}

Clearly, we see that $\rho^{BC}$ is independent of the secret $|\psi\rangle$. In other words, it is independent of the information parameter $\alpha$.

To answer (b), we consider the entire initial system as $\frac{1}{\sqrt{2}} (\alpha|0\rangle+\beta|1\rangle)(|000\rangle+|111\rangle)$. After Alice's and Bob's measurement the density operator of the entire system becomes

\begin{eqnarray}
\begin{split}
\rho_1 & = \frac{1}{8}\{|\phi^+\rangle(|+\rangle(\alpha|0\rangle+\beta|1\rangle) + |-\rangle(\alpha|0\rangle - \beta|1\rangle) ) \nonumber\\&\quad \quad
+ |\phi^-\rangle(|+\rangle(\alpha|0\rangle-\beta|1\rangle) + |-\rangle(\alpha|0\rangle + \beta|1\rangle) ) \nonumber\\&\quad \quad
+ |\psi^+\rangle(|+\rangle(\alpha|1\rangle+\beta|0\rangle) - |-\rangle(\alpha|1\rangle - \beta|0\rangle) ) \nonumber\\&\quad \quad
+ |\psi^-\rangle(|+\rangle(\alpha|1\rangle-\beta|0\rangle) - |-\rangle(\alpha|1\rangle + \beta|0\rangle) )\}\nonumber\\&\quad \quad \{\langle\phi^+|(\langle+|(\alpha\langle0|+\beta\langle1|) + \langle-|(\alpha\langle0| - \beta\langle1|) ) \nonumber\\&\quad \quad
+ \langle\phi^-|(\langle+|(\alpha\langle0|-\beta\langle1|) + \langle-|(\alpha\langle0| + \beta\langle1|) ) \nonumber\\&\quad \quad 
+\langle\psi^+|(\langle+|(\alpha\langle1|+\beta\langle0|) - \langle-|(\alpha\langle1| - \beta\langle0|) )  \nonumber\\&\quad \quad 
+ \langle\psi^-|(\langle+|(\alpha\langle1|-\beta\langle0|) - \langle-|(\alpha\langle1| + \beta\langle0|) )\}.
\end{split}
\end{eqnarray}

Tracing out Alice's and Bob's three-qubit system, the reduced density operator of Charlie's system is given by,

\begin{eqnarray}
\begin{split}
\rho_1^{C} & = \frac{1}{8}\{4(|\alpha|^2 +|\beta|^2)|0\rangle\langle0| + 4(|\alpha|^2 + |\beta|^2)|1\rangle\langle1|\}\nonumber\\ & = \frac{1}{2}\{|0\rangle\langle0| + |1\rangle\langle1|\} = \frac{I}{2}.
\end{split}
\end{eqnarray}

To answer (c), let us suppose Bob shares an additional resource $\frac{1}{\sqrt{2}}(|00\rangle+|11\rangle)$ with Alice. In this situation consider the entire initial system as $\frac{1}{2} (\alpha|0\rangle+\beta|1\rangle)_{A_0}(|000\rangle+|111\rangle)_{A_1B_1C_1}(|00\rangle+|11\rangle)_{A_2B_2}$. After Alice's, Bob's and Charlie's respective measurements of the revocation protocol the density operator of the entire system becomes $\rho_2 = \frac{1}{32}|Q\rangle\langle Q|$ where,

\begin{eqnarray}
\begin{split}
& |Q\rangle_{A_0A_1A_2B_1B_2C_1}\\ & =\{|\phi^+\rangle|s^+\rangle[|\phi^+\rangle|+\rangle +|\phi^-\rangle|-\rangle] + |\phi^+\rangle|s^-\rangle[|\phi^-\rangle|-\rangle +|\phi^-\rangle|+\rangle]\\&   
+ |\phi^+\rangle|r^+\rangle[|\psi^+\rangle|+\rangle -|\psi^-\rangle|-\rangle] + |\phi^+\rangle|r^-\rangle[|\psi^+\rangle|-\rangle -|\psi^-\rangle|+\rangle]\\&  
+ |\phi^-\rangle|s^+\rangle[|\phi^-\rangle|+\rangle -|\phi^+\rangle|-\rangle] + |\phi^-\rangle|s^-\rangle[|\phi^+\rangle|+\rangle +|\phi^-\rangle|-\rangle]\\& 
+ |\phi^-\rangle|r^+\rangle[|\psi^+\rangle|-\rangle -|\psi^-\rangle|+\rangle] + |\phi^-\rangle|r^-\rangle[|\psi^+\rangle|+\rangle -|\psi^-\rangle|-\rangle]\\& 
+ |\psi^+\rangle|s^+\rangle[|\psi^+\rangle|+\rangle +|\psi^-\rangle|-\rangle] - |\psi^+\rangle|s^-\rangle[|\psi^+\rangle|-\rangle +|\psi^-\rangle|+\rangle]\\& 
+ |\psi^+\rangle|r^+\rangle[|\phi^+\rangle|+\rangle +|\phi^-\rangle|-\rangle] - |\psi^+\rangle|r^-\rangle[|\phi^+\rangle|-\rangle +|\phi^-\rangle|+\rangle]\\& 
- |\psi^-\rangle|s^+\rangle[|\psi^+\rangle|+\rangle +|\psi^-\rangle|+\rangle] + |\psi^-\rangle|s^-\rangle[|\psi^+\rangle|+\rangle +|\psi^-\rangle|-\rangle]\\& 
- |\psi^-\rangle|r^+\rangle[|\phi^+\rangle|-\rangle +|\phi^-\rangle|+\rangle] + |\psi^-\rangle|r^-\rangle[|\phi^+\rangle|+\rangle +|\phi^-\rangle|-\rangle]\}\\& \nonumber
\end{split}
\end{eqnarray}

with,
\begin{align}
|s^+\rangle & = \alpha|0\rangle+\beta|1\rangle,\nonumber \\ |s^-\rangle & = \alpha|0\rangle-\beta|1\rangle,\nonumber \\ |r^+\rangle & = \alpha|1\rangle+\beta|0\rangle,\nonumber \\ |r^-\rangle & = \alpha|1\rangle-\beta|0\rangle.\nonumber
\end{align}
Tracing out Alice's system of three qubits corresponding to $\{A_0A_1A_2\}$, it is clear to see the reduced density operator of three-qubit system corresponding to $\{B_1B_2C_1\}$ is independent of the information parameter $\alpha$. \\ \\
To answer (d), let us suppose $R_i$ is the authorized receiver sending request to Charlie and sharing a Bell state $\frac{1}{\sqrt{2}}\{|00\rangle+|11\rangle\}_{CR_i}$ with him. In this situation, we consider the entire initial system as $\frac{1}{2} (\alpha|0\rangle+\beta|1\rangle)(|000\rangle+|111\rangle)(|00\rangle+|11\rangle)$. After Alice's, Bob's and Charlie's respective measurements the density operator of the entire system becomes $\rho_3 = \frac{1}{32}|R\rangle \langle R|$ where

\begin{eqnarray}
\begin{split}
|R\rangle & =\{|\phi^+\rangle|+\rangle[|\phi^+\rangle|s^+\rangle +|\phi^-\rangle|s^-\rangle + |\psi^+\rangle|r^+\rangle +|\psi^-\rangle|r^-\rangle ] \\& \quad 
 +|\phi^+\rangle|-\rangle[|\phi^+\rangle|s^-\rangle +|\phi^-\rangle|s^+\rangle + |\psi^+\rangle|r^-\rangle +|\psi^-\rangle|r^+\rangle ] \\& \quad 
 +|\phi^-\rangle|+\rangle[|\phi^+\rangle|s^-\rangle +|\phi^-\rangle|s^+\rangle + |\psi^+\rangle|r^-\rangle +|\psi^-\rangle|r^+\rangle] \\& \quad 
 +|\phi^-\rangle|-\rangle[|\phi^+\rangle|s^+\rangle +|\phi^-\rangle|s^-\rangle + |\psi^+\rangle|r^+\rangle +|\psi^-\rangle|r^-\rangle] \\& \quad 
 +|\psi^+\rangle|+\rangle[|\phi^+\rangle|r^+\rangle -|\phi^-\rangle|r^-\rangle + |\psi^+\rangle|s^+\rangle -|\psi^-\rangle|s^-\rangle] \\& \quad 
 -|\psi^+\rangle|-\rangle[|\phi^+\rangle|r^-\rangle -|\phi^-\rangle|r^+\rangle + |\psi^+\rangle|s^-\rangle -|\psi^-\rangle|s^+\rangle] \\& \quad 
 +|\psi^-\rangle|+\rangle[|\phi^+\rangle|r^-\rangle -|\phi^-\rangle|r^+\rangle + |\psi^+\rangle|s^-\rangle -|\psi^-\rangle|s^+\rangle] \\& \quad
 -|\psi^-\rangle|-\rangle[|\phi^+\rangle|r^+\rangle -|\phi^-\rangle|r^-\rangle + |\psi^+\rangle|s^+\rangle -|\psi^-\rangle|s^-\rangle]\}, \nonumber
\end{split}
\end{eqnarray}

with,
\begin{align}
|s^+\rangle & = \alpha|0\rangle+\beta|1\rangle,\nonumber \\ |s^-\rangle & = \alpha|0\rangle-\beta|1\rangle,\nonumber \\ |r^+\rangle & = \alpha|1\rangle+\beta|0\rangle,\nonumber \\ |r^-\rangle & = \alpha|1\rangle-\beta|0\rangle.\nonumber
\end{align}
Tracing out Alice, Bob and Charlie's five-qubit system, the reduced density operator of $R_i$'s system is again,

\begin{eqnarray}
\begin{split}
\rho_3^{R_i} & = \frac{1}{32}\{16(|\alpha|^2 +|\beta|^2)|0\rangle\langle0| + 16(|\alpha|^2 + |\beta|^2)|1\rangle\langle1|\}\nonumber\\ 
& = \frac{1}{2}\{|0\rangle\langle0| + |1\rangle\langle1|\} = \frac{I}{2}.
\end{split}
\end{eqnarray}

So here we find that in each of these steps, the reduced density matrix of the information seeker is independent of the information parameter $\alpha$. This clearly indicates that with semi-honest participants at every step our protocol is secure.\\ \\

\noindent\textit{\bf Concluding remarks and Outlook:}\\ \\
This paper addresses the problem of {\em revocable}\/ quantum secret sharing. The ability to revoke a quantum shared secret has implications for quantum routing (including backtracking ) in shared domain. An interesting consequence of the above is that critical/private information $\ket{S}$ can be {\em q-mailed} across public \textsc{qnet}s, first by secret sharing $\ket{S}$ and then routing $\ket{S}$ (in the shared domain) to the desired receiver. We have assumed the resources to be pure entangled states; however, working out with resources being mixed entangled states still remains an open question. Another direction can be of sharing multi-partite entangled qubits and routing them to the desired receiver. Yet another direction would be to use the techniques of \cite{karlsson} to thwart attacks beyond the semi-honest adversary \cite{Qin}. Recently, cryptanalysis of other QSS protocols has been discussed in \cite{Wang, Wangli}.\\ \\

\noindent\textit{\bf Acknowledgment:} This work is done at  Center for Security, Theory and Algorithmic Research (CSTAR), IIIT, Hyderabad, and S Sazim gratefully acknowledge their hospitality. We acknowledge Prof. P. Agrawal for useful discussions. 

\newpage
\begin{appendix}
\begin{widetext}

\noindent\textit{\bf Appendix 1:}\\
Consider a 3-qubit GHZ state $\frac{1}{\sqrt{2}}\{|000\rangle$+$|111\rangle\}$ among Alice, Bob and Charlie and let $|\psi\rangle = \alpha|0\rangle+\beta|1\rangle$ be the message with Alice.
\begin{eqnarray}
&&|\psi\rangle\otimes\frac{1}{\sqrt{2}}\{|000\rangle+|111\rangle\}{}\nonumber\\&& 
= \{\alpha|0\rangle+\beta|1\rangle\}\otimes\frac{1}{\sqrt{2}}\{|000\rangle+|111\rangle\} {}\nonumber\\&& 
= \frac{1}{\sqrt{2}}\{\alpha|0000\rangle+\alpha|0111\rangle+\beta|1000\rangle+\beta|1111\rangle\} {}\nonumber\\&& 
= \frac{1}{\sqrt{2}}\{|00\rangle\alpha|00\rangle+|01\rangle\alpha|11\rangle+|10\rangle\beta|00\rangle+|11\rangle\beta|11\rangle\} {}\nonumber\\&&
= \frac{1}{2}\{[|\phi^+\rangle + |\phi^-\rangle]\alpha|00\rangle+[|\psi^+\rangle + |\psi^-\rangle]\alpha|11\rangle+[|\psi^+\rangle - |\psi^-\rangle]\beta|00\rangle+[|\phi^+\rangle - |\phi^-\rangle]\beta|11\rangle\}{}\nonumber\\&&
= \frac{1}{2}\{|\phi^+\rangle[\alpha|00\rangle +\beta|11\rangle] + |\phi^-\rangle[\alpha|00\rangle-\beta|11\rangle]+|\psi^+\rangle[\alpha|11\rangle+\beta|00\rangle]+|\psi^-\rangle[\alpha|11\rangle-\beta|00\rangle]\}
\end{eqnarray}

\noindent\textit{\bf Appendix 2:}\\
Suppose Alice and Bob share a bell state $\frac{1}{\sqrt{2}}\{|00\rangle+|11\rangle\}_{AB}$ 
and the secret is already being shared between Bob and Charlie is $\{\alpha|00\rangle+\beta|11\rangle\}_{BC}$.
\begin{eqnarray}
&&\frac{1}{\sqrt{2}}\{|00\rangle+|11\rangle\}_{AB}\otimes\{\alpha|00\rangle+\beta|11\rangle\}_{BC}{}\nonumber\\&&  
= \frac{1}{\sqrt{2}}\{\alpha|0\rangle|00\rangle|0\rangle+\beta|0\rangle|01\rangle|1\rangle+\alpha|1\rangle|10\rangle|0\rangle+\beta|1\rangle|11\rangle|1\rangle\}_{ABBC} {}\nonumber\\&&
= \frac{1}{2\sqrt{2}}\{
\alpha|0\rangle[|\phi^+\rangle+|\phi^-\rangle][|+\rangle+|-\rangle] 
 + \beta|0\rangle[|\psi^+\rangle+|\psi^-\rangle][|+\rangle-|-\rangle] {}\nonumber\\&&
+ \alpha|1\rangle[|\psi^+\rangle-|\psi^-\rangle][|+\rangle+|-\rangle]
 + \beta|1\rangle[|\phi^+\rangle-|\phi^-\rangle][|+\rangle-|-\rangle]
\} {}\nonumber\\&&
=\frac{1}{2\sqrt{2}}\{
[\alpha|0\rangle+\beta|1\rangle]|\phi^+\rangle|+\rangle
 + [\alpha|0\rangle-\beta|1\rangle]|\phi^-\rangle|+\rangle 
 + [\alpha|0\rangle-\beta|1\rangle]|\phi^+\rangle|-\rangle{}\nonumber\\&&
 + [\alpha|0\rangle+\beta|1\rangle]|\phi^-\rangle|-\rangle 
 + [\beta|0\rangle+\alpha|1\rangle]|\psi^+\rangle|+\rangle
 + [\beta|0\rangle-\alpha|1\rangle]|\psi^-\rangle|+\rangle{}\nonumber\\&& 
 + [\alpha|1\rangle-\beta|0\rangle]|\psi^+\rangle|-\rangle
 - [\alpha|1\rangle+\beta|0\rangle]|\psi^-\rangle|-\rangle
\}
\end{eqnarray}

\noindent\textit{\bf Appendix 3:}\\
Suppose $R_i$ is the authorized receiver sending request to Charlie and sharing a bell state $\frac{1}{\sqrt{2}}\{|00\rangle+|11\rangle\}_{CR_i}$ with charlie. Suppose $\{\alpha|00\rangle+\beta|11\rangle\}_{BC}$ is shared among Bob and Charlie.
\begin{eqnarray}
&&\{\alpha|00\rangle+\beta|11\rangle\}_{BC}\otimes\frac{1}{\sqrt{2}}\{|00\rangle+|11\rangle\}_{CR} {}\nonumber\\&&
=\frac{1}{\sqrt{2}}\{\alpha|0\rangle|00\rangle|0\rangle+
\alpha|0\rangle|01\rangle|1\rangle+\beta|1\rangle|10\rangle|0\rangle+\beta|1\rangle|11\rangle|1\rangle\}_{BCCR} {}\nonumber\\&&
= \frac{1}{2\sqrt{2}}\{
[|+\rangle+|-\rangle][|\phi^+\rangle+|\phi^-\rangle]\alpha|0\rangle 
+ [|+\rangle+|-\rangle][|\psi^+\rangle+|\psi^-\rangle]\alpha|1\rangle {}\nonumber\\&&
+ [|+\rangle-|-\rangle][|\psi^+\rangle-|\psi^-\rangle]\beta|0\rangle 
+ [|+\rangle-|-\rangle][|\phi^+\rangle-|\phi^-\rangle]\beta|1\rangle
\} {}\nonumber\\&&
=\frac{1}{2\sqrt{2}}\{
|+\rangle|\phi^+\rangle[\alpha|0\rangle+\beta|1\rangle]
 + |+\rangle|\phi^-\rangle[\alpha|0\rangle-\beta|1\rangle] {}\nonumber\\&&
 + |-\rangle|\phi^+\rangle[\alpha|0\rangle-\beta|1\rangle]
 + |-\rangle|\phi^-\rangle[\alpha|0\rangle+\beta|1\rangle] 
 + |+\rangle|\psi^+\rangle[\alpha|1\rangle+\beta|0\rangle]{}\nonumber\\&&
 + |+\rangle|\psi^-\rangle[\alpha|1\rangle-\beta|0\rangle]
 + |-\rangle|\psi^+\rangle[\alpha|1\rangle-\beta|0\rangle]
 + |-\rangle|\psi^-\rangle[\alpha|1\rangle+\beta|0\rangle]
\}
\end{eqnarray}

\noindent\textit{\bf Appendix 4:}\\
Suppose $(A_1, B_1)$ has the shared secret $\{\alpha|00\rangle+\beta|11\rangle\}_{A_{1s}B_{1s}}$ as a result of bell measurement at the sender $(S)$. Assume the pairs $(A_1, A_2), (B_1, B_2)$ and $(B_2, R)$ share Bell states $\frac{1}{\sqrt{2}}\{|00\rangle+|11\rangle\}_{A_{1r}A_{2r}}, \frac{1}{\sqrt{2}}\{|00\rangle+|11\rangle\}_{B_{1r}B_{2r}}$ and $\frac{1}{\sqrt{2}}\{|00\rangle+|11\rangle\}_{B_{2r}R}$ respectively. The following is a sequence of measurements that transfers the secret in shared form from $(A_1, A_2)$ to $(B_1, B_2)$ so that $R$ can be able to reconstruct it. 
\begin{enumerate}
\item Bell Measurement at $A_1$ on qubits $A_{1s}$ and $A_{1r}$.
\begin{eqnarray}
&&\{\alpha|00\rangle+\beta|11\rangle\}_{A_{1s}B_{1s}}\otimes\frac{1}{\sqrt{2}}\{|00\rangle+|11\rangle\}_{A_{1r}A_{2r}} {}\nonumber\\&&
=\frac{1}{\sqrt{2}}\{\alpha|00\rangle|00\rangle+
\alpha|01\rangle|01\rangle+\beta|10\rangle|10\rangle+\beta|11\rangle|11\rangle\}_{A_{1s}A_{1r}B_{1s}A_{2r}} {}\nonumber\\&&
= \frac{1}{2}\{\alpha[|\phi^+\rangle + |\phi^-\rangle]|00\rangle+\alpha[|\psi^+\rangle + |\psi^-\rangle]|01\rangle+\beta[|\psi^+\rangle - |\psi^-\rangle]|10\rangle+\beta[|\phi^+\rangle - |\phi^-\rangle]|11\rangle\}_{A_{1s}A_{1r}B_{1s}A_{2r}} {}\nonumber\\&&
= \frac{1}{2}\{|\phi^+\rangle[\alpha|00\rangle +\beta|11\rangle] + |\phi^-\rangle[\alpha|00\rangle-\beta|11\rangle]+|\psi^+\rangle[\alpha|01\rangle+\beta|10\rangle]+|\psi^-\rangle[\alpha|01\rangle-\beta|10\rangle]\}_{A_{1s}A_{1r}B_{1s}A_{2r}}
\end{eqnarray}
Suppose $|\psi^+\rangle$ is the outcome of this measurement, the new state of the secret will be $\{\alpha|01\rangle+\beta|10\rangle\}_{B_{1s}A_{2s}}$ with $(B_1,A_2)$. This outcome will be sent to $S$ classically to keep track of the present state of the secret.
\item Bell Measurement at $B_1$ on qubits $B_{1s}$ and $B_{1r}$.
\begin{eqnarray}
&&\{\alpha|01\rangle+\beta|10\rangle\}_{B_{1s}A_{2s}}\otimes\frac{1}{\sqrt{2}}\{|00\rangle+|11\rangle\}_{B_{1r}B_{2r}} {}\nonumber\\&&
=\frac{1}{\sqrt{2}}\{\alpha|00\rangle|10\rangle+
\alpha|01\rangle|11\rangle+\beta|10\rangle|00\rangle+\beta|11\rangle|01\rangle\}_{B_{1s}B_{1r}A_{2s}B_{2r}} {}\nonumber\\&&
= \frac{1}{2}\{\alpha[|\phi^+\rangle + |\phi^-\rangle]|10\rangle+\alpha[|\psi^+\rangle + |\psi^-\rangle]|11\rangle+\beta[|\psi^+\rangle - |\psi^-\rangle]|00\rangle+\beta[|\phi^+\rangle - |\phi^-\rangle]|01\rangle\}_{B_{1s}B_{1r}A_{2s}B_{2r}} {}\nonumber\\&&
= \frac{1}{2}\{|\phi^+\rangle[\alpha|10\rangle +\beta|01\rangle] + |\phi^-\rangle[\alpha|10\rangle-\beta|01\rangle]+|\psi^+\rangle[\alpha|11\rangle+\beta|00\rangle]+|\psi^-\rangle[\alpha|11\rangle-\beta|00\rangle]\}_{B_{1s}B_{1r}A_{2s}B_{2r}}
\end{eqnarray}
Suppose $|\phi^-\rangle$ is the outcome of this measurement, the new state of the secret will be $\{\alpha|10\rangle-\beta|01\rangle\}_{A_{2s}B_{2s}}$ with $(A_2,B_2)$. This outcome will be sent to $S$ classically to keep track of the present state of the secret.

\end{enumerate}

\end{widetext}
\end{appendix}
\end{document}